\documentclass[12pt]{article}
\usepackage{amsfonts}
\usepackage{amssymb}
\usepackage{graphicx}
\usepackage{color}
\usepackage{amsmath}
\usepackage{amsthm}
\usepackage{setspace}
\usepackage{fullpage}
\usepackage{natbib}

\usepackage{setspace}
\begin{document}
\renewcommand*{\thefootnote}{\fnsymbol{footnote}}

\title{\vspace{-1cm} Why Indexing Works}

\author{J. B. Heaton \footnote{Conjecture LLC, jb@conjecturellc.com} \and N. G. Polson \footnote {Booth School of Business,  
University of Chicago, ngp@chicagobooth.edu} \and J. H. Witte \footnote{Department of Mathematics, University College London, and Conjecture LLC, jhw@conjecturellc.com}}

\date{\small{First Draft: October 2015\\This Draft: May 2017}}

\maketitle


\begin{abstract}
\bigskip

We develop a simple stock selection model to explain why active equity managers tend to underperform a benchmark index. We motivate our model with the empirical observation that the best performing stocks in a broad market index often perform much better than the other stocks in the index. Randomly selecting a subset of securities from the index may dramatically increase the chance of underperforming the index. The relative likelihood of underperformance by investors choosing active management likely is much more important than the loss those same investors take due to the higher fees of active management relative to passive index investing. Thus, active management may be even more challenging than previously believed, and the stakes for finding the best active managers may be larger than previously assumed.\\

\end{abstract} \vspace{-.6cm}

\pagenumbering{gobble}
\clearpage

\newpage
\singlespacing
\pagenumbering{arabic}

\section{Introduction}\label{sec:intro} 

\renewcommand*{\thefootnote}{\arabic{footnote}}

The tendency of active equity managers  to underperform a passive benchmark index (\textit{e.g.}, Lakonishok, Shleifer, and Vishny (1992), Gruber (1996)) is something of a mystery. It is one thing for active equity managers to fail \textit{to beat} the benchmark index, since that may imply only a lack of skill to do better than random selection. It is quite another to find that active equity managers very often fail \textit{to keep up with} the benchmark index, since that implies that active equity managers are doing something that systematically leads to underperformance. 

We develop a simple stock selection model that builds on the underemphasized empirical fact that the best performing stocks in a broad index often perform much better than the other stocks in the index, so that average index returns depend heavily on a relatively small set of winners (\textit{e.g.}, J.P. Morgan (2014)).  In our model, randomly selecting a small subset of securities from an index maximizes the chance of outperforming the index - the allure of active equity management - but it also maximizes the chance of underperforming the index, with the chance of underperformance being larger than the chance of overperformance.  To illustrate the idea, consider an index of five securities, four of which (though it is unknown which) will return 10\% over the relevant period and one of which will return 50\%. Suppose that active managers choose portfolios of one or two securities and that they equally-weight each investment. There are 15 possible one or two security ``portfolios."  Of these 15, 10 will earn returns of 10\%, because they will include only the 10\% securities. Just 5 of the 15 portfolios will include the 50\% winner, earning 30\% if part of a two security portfolio and 50\% if it is the single security in a one security portfolio.  The mean average return for all possible actively-managed portfolios will be 18\%, while the median portfolio of all possible one- and two-stock portfolios will earn 10\%. The equally-weighted index of all 5 securities will earn 18\%. Thus, in this example, the average active-management return will be the same as the index (see Sharpe (1991)), but two-thirds of the actively-managed portfolios will underperform the index because they will omit the 50\% winner.

In this example, it is a large positive skewness in returns that creates a problem for active management, illustrated here as the selection of one or two securities. The non-symmetric shape of the distribution of returns means that random selection - which we might think of as a plausible lower bound on the quality of active management - will deliver a median return that is  worse than the average of the full index of the securities.

In reality, the histrogram of returns to the securities in an index will change year-to-year. Our model presents this as a problem of skewness, but our point is more general. One reason indexing ``works" so well is that, on average it seems, active management faces a higher hurdle than previously recognized. Missing (or underweighting) the securities that significantly outperform other securities is a strong headwind for an active manager to overcome. This view of the active-passive problem helps us understand the mystery of how so many smart people, with enormous financial and informational resources, systematically do such a poor job investing money. 

Our paper continues as follows. In Section 2, we develop our simple stock selection model and comment on relationships with sets of empirical data. In Section 3, we present a Monte Carlo simulation of our model. Section 4 concludes.

\section{A Simple Model of Stock Selection from an Index}

We consider a benchmark index that contains  $N$ stocks $S^i$, $1\leq i\leq N$. Let the dynamics of stock $S^i$ over time $t\in [0,T]$ be given by a geometric Brownian motion
\begin{equation*}
\frac{dS^i_{t+1}}{S^i_t} = \mu_i\, dt + \sigma\,dW_t,
\end{equation*}
where for simplicity we consider the volatility $\sigma>0$ to be constant for all stocks. We assume that stock drifts are distributed $\mu_i\sim N(\hat{\mu},\hat{\sigma}^2)$, which generates a small number of extreme winners, a small number of extreme losers, and a large number of stocks with drifts centered around $\hat{\mu}$ with standard deviation $\hat{\sigma}>0$. While our model implies unpriced covariance among securities and a lack of learning, much theory and evidence suggests that the learning problem is too difficult over the lifetimes of most investors to pay much attention to that modeling limitation (\textit{e.g.}, Merton (1980), Jobert, Platania, and Rogers (2006)).\footnote{In one study of stock market fluctuations,  Barsky and DeLong (1993) discuss the problem of estimating a particular parameter for an assumed dividend process, noting that a Bayesian updater might not be shifted significantly from his prior after 120 years of data and that ``[e]ven if we were lucky and could precisely estimate [the parameter], no investor in 1870 or 1929-lacking the data that we possess-had any chance of doing so."} In any case, our main goal is to generate a set of returns that - like we often see empirically - have a set of winners that significantly outperforms other members of the index.

For simplicity, we assume that individual stocks maintain their drift $\mu_i$ over the time period $t\in [0,T]$. We also assume that individual stocks have a starting value $S^i_0=1$ for all stocks.

If at time $t=0$ we pick a stock $S^i_0$ at random, then our time $T$ value follows
\begin{align*}
S^i_T\sim e^{\hat{\mu}T-\frac{1}{2}\sigma^2T + \sqrt{\sigma^2T+\hat{\sigma}^2T^2}Z}, 
\end{align*}
where $Z\sim N(0,1)$, provided we assume $\mu_i$ and $W_T$ are independent.

We define an index return by the equally weighted portfolio
\begin{equation*}
I^N_t=\frac{1}{N}\sum^N_{i=1}S^i_t,
\end{equation*}
which corresponds to a capital weighted index of $N$ stocks.

Two observations are apparent. First, the cumulative return of a stock picked randomly at time $t=0$ follows a log-$N(\hat{\mu} T -\frac{1}{2}\sigma^2T, \sigma^2T+\hat{\sigma}^2T^2)$ distribution. The variance component $\hat{\sigma}^2T^2$, which indicates the over-proportional profit a continuously compounded winner will bring relative to the loss incurred by a loser. That is, the distribution is heavily positively skewed with a mean of $e^{\hat{\mu}T+\frac{1}{2}\hat{\sigma}^2T^2}$. Second, the median of the stock distribution is given by $e^{\hat{\mu} T-\frac{1}{2}\sigma^2 T}$, so that over time $T$ more than half of all stocks in the index will underperform the index return  $I^N_T$ by a factor of $e^{\frac{1}{2}\sigma^2T+\frac{1}{2}\hat{\sigma}^2T^2}$. 

\begin{figure}[th!]
\centering
\includegraphics[height=0.5\textwidth, width=1\textwidth]{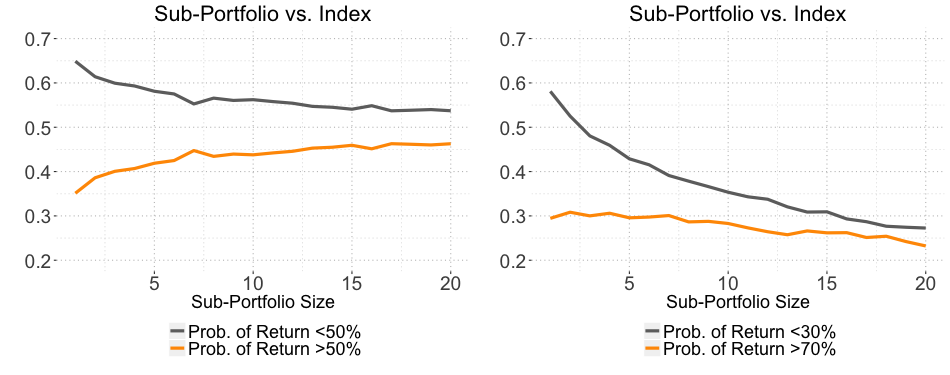}
\caption{\footnotesize{\emph{On the left, overlapping frequencies of over- and underperformance relative to index average return of 50\%. On the right, overlapping frequencies of 20\% over- and underperformance relative to index average return 50\%. While random selection of small sub-portfolios has the greatest probability of getting overperformance, it also risks a relatively high probability of underperformance. The risk of substantial index underperformance always dominates the chance of substantial index outperformance and is greatest for small portfolios.}}}
\label{fig:ModelQuantiles}
\end{figure}

\begin{figure}[th!]
\centering
\includegraphics[height=0.5\textwidth, width=0.8\textwidth]{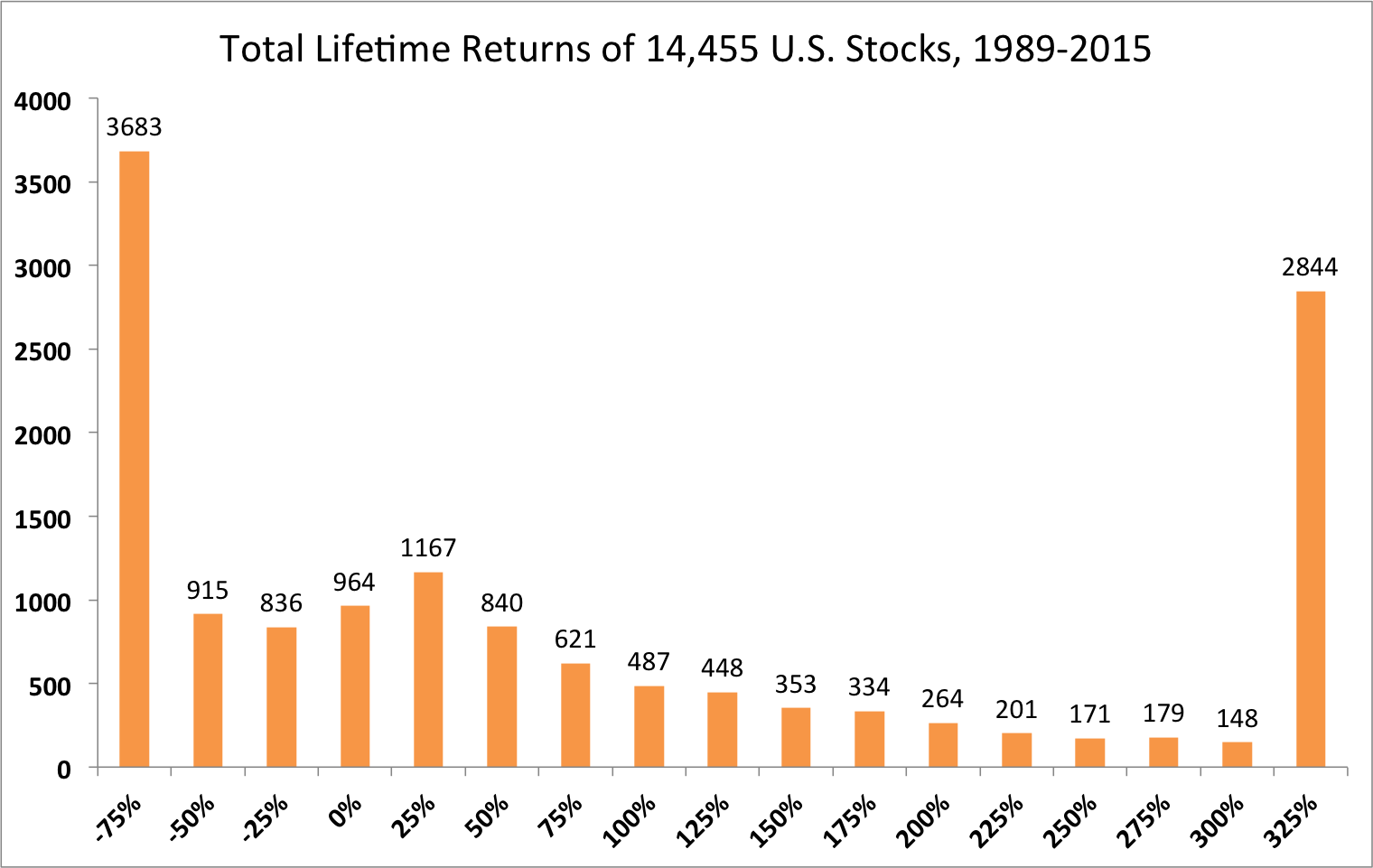}
\caption{\footnotesize{\emph{This chart can be found on http://awealthofcommonsense.com/2016/05/the-sp-500-is-the-worlds-largest-momentum-strategy/. We see that 40\% of all stocks generated no return (while the S\&P500 was up almost 1200\% over the same period).}}}
\label{fig:LifetimeReturns}
\end{figure}

Another interpretation is that a geometric Brownian motion $$ S_t = S_0 \exp \left ( \mu t - \frac{1}{2} \sigma^2 t \right )  \exp \left ( \sigma B_t \right ) $$ has first moment $ \mathbb{E}(S_t ) = S_0 e^{\mu t} $ and mode $ S_0 e^{\mu t- \frac{1}{2} \sigma^2 t }$ . By the strong law of large numbers, for any $ \epsilon > 0 $, $ \mathbb{P} \left ( - \epsilon t < B_t < \epsilon t \right ) \rightarrow 1 $
as $ t \rightarrow \infty $. Therefore, for large enough values of $t$,
$$
e^{ - \epsilon t} < S_0 e^{-\gamma t} < e^{\epsilon t} \; \; {\rm where} \; \; \gamma = \mu - \frac{1}{2} \sigma^2.
$$
Counter intuitively, the realised stock value is no where near its mean, as the growth rate $  \gamma = \mu - \frac{1}{2} \sigma^2 \ll \mu $. Clearly, it is the median that governs the long run.
For example, a portfolio $(\omega , 1 - \omega ) $ of stock and risk-free rate will have an expected return $ \mu_\omega = r + \omega ( \mu - r ) $ and growth rate and volatility
$
\gamma_\omega =r + \omega ( \mu -  r ) \frac{1}{2} \omega^2\sigma^2 \; \; {\rm and} \; \; \sigma_\omega = \omega \sigma.
$


Another interesting observation is that of Jobert et al. (2006) who explain why it is so rare to achieve the same performance as the mean. Namely, if you observe daily prices for a stock with annual return and volatility of $20$\%, then you need about 11 years of data to provide a confidence interval of $ \pm 1 $\% around the estimated volatility of the assumed underlying stochastic process. Conversely, you require about 1550 years of data to estimate the return with the same precision.

\vspace{0.1in}

On the empirical side, it is worth noting just how astonishing the wealth generation of indexing with only a very small proportion of winners has been for investors.
For example, Bessembinder (2017) analyses the $26,000$ stocks that have entered the CRSP database from 1926 until 2015. He finds that $ 58 $\% of common stocks have under-performed the T-bill rate over their full lifetime. Moreover, the entire gain in
the U.S. stock-market since 1926 is attributable to only $4$\% of the stocks. The top $86$ stocks have created a $50$\% lion-share of the 
total $ \$ 32$ trillion dollars achieved. These effects do not seem to be disappearing; for example,
Figure \ref{fig:LifetimeReturns} shows a similar effect for the period 1989-2015.  For example, the skewness in individual stock winners such as  Amazon which has returned $ 35,000 $\% from 1999 versus 
$181$\% for the S\&P500 index is dramatic. Again, as in the full sample, more than $50$\% of the stocks in this period have under-performed cash. 

\section{Monte Carlo Simulation}

To illustrate our effect, we provide a simple Monte Carlo simulation.
We assume a median index return of $10\%$ and an expected index return of $50\%$ over the considered period $T$. We take $\sigma=20\%$ as a generic annual stock volatility. We choose $T=5$ (five years), $\hat{\mu}=(\log 1.1 + \frac{1}{2}0.2^2\cdot 5)/5\approx 4\%$, and $\hat{\sigma}=\sqrt{2 \log 1.5-0.04\cdot 5\cdot 2}/5\approx 13\%$. We show the frequency of exceeding or falling short of the expected five year, 500-stock index return $\mathbb{E}I^{N=500}_{T=5}-1\approx50\%$ when creating sub-portfolios of different sizes (each computed based on a Monte Carlo simulation with 10,000 samples). Figure \ref{fig:ModelQuantiles} left shows the frequency with which randomly selected portfolios of a given size overperform (5 year return greater than 50\%) and underperform (5 year return less than 50\%) the expected return for all 500 stocks.  Figure \ref{fig:ModelQuantiles} right shows the frequency with which randomly selected portfolios of a given size overperform (5 year return greater than 70\%) and underperform (5 year return less than 30\%) using more extreme thresholds for over- and underperformance.


The risk of substantial index underperformance always dominates the chance of substantial index outperformance, with the difference being greater the smaller the size of the selected sub-portfolios. It is far more likely that a randomly selected (small) subset of the 500 stocks will underperform than overperform, because average index performance depends on the inclusion of the extreme winners that often are missed in sub-portfolios. 

\section{Conclusion}

Researchers have focused on the costs of active management as being primarily the fees paid for active management (\textit{e.g.}, French 2008). Our model (which is but one way of looking at the problem) suggests that the much higher cost of active management may be the inherently high chance of underperformance that comes with attempts to select stocks, since stock selection itself increases the chance of underperformance relative to the chance of overperformance in many circumstances. To the extent that those allocating assets have assumed that the only cost of active investing above indexing is the cost of the active manager in fees, that assumption should be revisited. Active managers do not start out on an even playing field with passive investing. Rather, active managers must overcome an inherent disadvantage. The stakes for identifying the best active managers may be higher than previously thought.

Put another way, passive investing may have a larger head start on active investing than previously believed. When creating a portfolio combining passive and active strategies, independently of past performance, return estimation should be adjusted for the inherent statistical disadvantage of the active manager combined with their higher fees.  

\vspace{0.1in}



\footnotesize

\subsection*{References}\vspace{-0.5cm}

\singlespacing

\noindent Barsky, R. B. and J. B. DeLong. (1993). Why does the stock market fluctuate? \textit{The Quarterly Journal of Economics}, 108, 291-311.\medskip

\noindent Bessembinder, H. (2017). Do Stocks Outperform Treasury Bills?, https://ssrn.com/abstract=2900447. \medskip

\noindent French, K.R. (2008). Presidential Address: The Cost of Active Investing. \textit{Journal of Finance}, 63(4), 1537-73.\medskip

\noindent Gruber, M.J. (2008). Another Puzzle: The Growth in Actively Managed Mutual Funds. \textit{Journal of Finance}, 51(3), 783-810.\medskip

\noindent Ikenberry, D.L., R.L. Shockley, and K.L. Womack. (1992). Why Active Managers Often Underperform the S\&P500: The Impact of Size and Skewness, \textit{Journal of Private Portfolio Management}, 1(1), 13-26. \medskip

\noindent J.P. Morgan. (2014). Eye on the Market, Special Edition: The Agony \& the Ecstasy: The Risks and Rewards of a Concentrated Stock Position. \medskip

\noindent Jobert, A., A. Platania, and L. C. G. Rogers. (2006). A Bayesian solution to the equity premium puzzle. Unpublished paper, Statistical Laboratory, University of Cambridge.\medskip

\noindent Lakonishok, J., A. Shleifer and R. Vishny. (1992). 
The Structure and Performance of the Money Management Industry. \textit{Brookings Papers on Economic Activity: 
Microeconomics}, 339-391.\medskip

\noindent Merton, R. C. (1980). On Estimating the Expected Return on the Market. \textit{Journal of Financial
Economics}, 8, 323-361.\medskip

\noindent Sharpe, W.F.(1991). The Arithmetic of Active Management. \textit{Financial Analysts Journal}, 47(1), 7-9. \medskip

\singlespacing

\end{document}